\documentclass[twocolumn,secnumarabic,amssymb, nobibnotes, aps, prd]{revtex4-2}

\newcommand{\env}[1]{\texttt{#1}}
\usepackage{graphicx, bm}
\usepackage{float}
\begin{document}

\title{Tangent Vector Variational Quantum Eigensolver: Robust Variational Quantum Eigensolver against the inaccuracy of derivative}

\author{Hikaru Wakaura}%
\email[Quantscape: ]{
hikaruwakaura@gmail.com}
\affiliation{QuantScape Inc., 4-11-18, Manshon-Shimizudai, Meguro, Tokyo, 153-0064, Japan}

\author{Andriyan B. Suksmono}

\affiliation{ Institut Teknologi Bandung, Jl. Ganesha No.10, Bandung, Jawa Barat, Indonesia}
\email[Bandung Institute of Technology: ]{suksmono@stei.itb.ac.id}

\date{December 2018}%

\begin{abstract}
Since the release of cloud quantum computers in 2016, research of the Noisy-Intermediate-Scale-Quantum (NISQ) algorithm has accelerated over the world. Variational Quantum Algorithms (VQAs) are NISQ algorithms that optimize the evaluation functions using both quantum computers and classical optimizers. Especially, Variational Quantum Eigensolver (VQE) optimizes the energy of quantum systems. Besides, some advantageous methods are proposed to derive the excited states and their energies accurately. However, present methods can't derive the excited energies accurately. Therefore, we propose the novel VQE method that minimizes the tangent vector of parameter landscape. We call this Tangent-Vector VQE (TVVQE) that is used after the VQE method. We demonstrate the calculation of energy levels on hydrogen molecule, rectangular Hubbard Hamiltonian, the Hamiltonian of lithium hydride molecule, and the Hamiltonian of beryllium hydride molecule and reveal that TVVQE has a potential application for calculation of the ground and excited energy levels more accurately than other VQE methods. This algorithm has the potential to derive the excited energies of fault-tolerant quantum algorithms on mid-scale NISQ computers that have less than 50 qubits.
\newline

\end{abstract}

\maketitle

\section{Introduction}\label{1}
Since introduced by Aspuru-Guzik et al. in 2011 \cite{doi:10.1146/annurev-physchem-032210-103512}, the Variational Quantum Eigensolver (VQE) has been investigated and improved by various groups. Cloud quantum computing with 5 qubits by IBM initiated the movement to benchmark the calculation of the VQE method on various molecules \cite{PhysRevX.8.011021, 2019arXiv190210171N, 2017Natur.549..242K} and quantum systems, which eventually accelerate its further development. Today, various kinds of VQE methods are brought out to the world. For example, Subspace-Search VQE (SSVQE) \cite{PhysRevResearch.1.033062} can calculate the multiple energy levels at once and Multiscale-Contracted VQE (MCVQE) \cite{2019PhRvL.122w0401P} calculates the ground and single electron excited states by diagonalizing configuration interaction state Hamiltonian. Adaptive VQE \cite{2019NatCo..10.3007G} and Deep-VQE \cite{2020arXiv200710917F} are also proposed. The essential procedure of VQE has been exploited for machine learning \cite{PhysRevA.98.032309}.

In parallel to this movement, quantum hardware has been improved concerning both the number of qubits and quantum volume. The fidelity of qubits skyrocketed last year. Both Honeywell \cite{2020PhRvR...2a3317B} and Ion-Q \cite{IonQ2020} updated the record of quantum volume twice and the record of this is 1024 achieved by Quantinuum(formally Honeywell)\cite{2021arXiv211003137L}.

Recently, the Institute of Science in China has achieved quantum supremacy by photonic quantum computer \cite{Zhong1460}. It will not be long before they will realize a Fault-Tolerant Quantum Computer (FTQC) \cite{Gambetta2017}. The FTQC devices require many error-correcting qubits for the given number of logical qubits. The FTQC with 100 logical qubits requires over 10000 physical qubits using topological surface code. Primitive FTQC devices that have less than 10 logical qubits will be used as large-scale Noisy-Intermediate-Scale-Quantum (NISQ) devices.
Hence, the VQE method will be used until FTQC with more than 100 logical qubits are released and it is required to be modified to improve the accuracy using the quantum resource of the FTQC.

Optimization of variables is performed by classical computers, thus, gradients and Hessians of trial energy are calculated by Parameter-shift-rule(PSR) regardless of the difference between the values derived analytically and PSR \cite{2019arXiv190608728P}. It is because   the gradient and Hessians cannot be derived by quantum computers fast enough. Gradients derived by PSR are less accurate than analytical gradients. This declines the accuracy of the VQE method. In the case not only energy level but also gradients are optimized, the accuracy of the results must be improved. Therefore, we propose an improved VQE method that optimizes the gradient of trial energy calculated analytically. We call this method Tangent-Vector VQE (TVVQE) method. We confirmed that this TVVQE has the potential to calculate the energy levels and states in high accuracy compared to other methods.

The rest of this paper is as follows. Section \ref{2} describes the detail of our method. Section \ref{3.25} compares the result of the calculation on energy levels of three types of quantum systems. In Section \ref{3.75}, we compare the result of the calculation on the ground, triplet, singlet, and doubly excited state derived by TVVQE to other well-known methods. Section \ref{4} concludes this work.

\section{Method}\label{2}
In the TVVQE method, we optimize the tangent vector of trial energy $\langle \Phi_j \mid H \mid \Phi_j \rangle$ for each variable instead of trial energy. The ordinary VQE is a hybrid method that calculates trial energy on quantum computers for variables and optimizes the variables in order to find the minimum of trial energy. The equation of the trial energy for Hamiltonian H is as follows,

\begin{equation}
E_j(\bm{\theta^j})=\langle \Phi_{ini} \mid U^\dagger(\bm{\theta^j})HU(\bm{\theta^j})\mid \Phi_{ini} \rangle\label{E},
\end{equation}

where $\mid \Phi_{ini} \rangle$ is initial state, $U(\bm{\theta^j})$ is the operator to make the given superposition state that includes the trotterized Hamiltonian and cluster terms correspond to $\bm{\theta^j}$, that is the variable vector of $j^{th}$ state, respectively. $U(\bm{\theta^j})$ . The depth of Hamiltonian and cluster terms are set to $n$, which in the following discussions is 2. It is expressed as $U(\bm{\theta^j})=\prod_l \prod_k exp(-i\theta_l^j c_k^lP_k^l)$ by the index of the variable $l$. $P_k^l$ indicates the $k^{th}$ Pauli opetrator transformed from $l^{th}$ term by Bravyi-Kitaev transformation \cite{doi:10.1021/acs.jctc.8b00450} \cite{2017arXiv171007629M}. We use Unitary Coupled Cluster(UCC) \cite{2018PhRvA..98b2322B} ansatz to calculate the derivatives. The function to be minimized in actual VQE method is a function that contains constraint terms \cite{doi:10.1021/acs.jctc.8b00943} $E^{const}_j(\bm{\theta^j})$ and deflation terms of Variational Quantum Deflation(VQD) \cite{2018arXiv180508138H} $E^{def}_j(\bm{\theta^j})$.The method to calculate the product between two states is SWAP-test \cite{2013PhRvA..87e2330G}, which is used to calculate the excited states. They are for calculation of excited states. The evaluation function of $j^{th}$ state is,

\begin{equation}
F(\bm{\theta^j})=E_j(\bm{\theta^j})+E^{const}_j(\bm{\theta^j})+E^{def}_j(\bm{\theta^j})\label{F}.
\end{equation}

The constraint and deflation terms are zero at a global minimum of evaluation function. TVVQE minimizes the norm of tangent vector instead of trial energy after optimizing eq.(\ref{F}). Derivatives in tangent vector are calculated by the method \cite{2017arXiv170102691R}. Hence, the derivative of variable $\theta_i$ is,

\begin{equation}
\partial E_j/\partial \theta_m^j = 2\sum_{k=0}c_k^mIm(\langle \Phi_{ini} \mid U^\dagger(\bm{\theta^j}) HV_k^m(\bm{\theta^j}) \mid \Phi_{ini} \rangle)\label{deri},
\end{equation}

which can be derived by Hadamard-test-like method. Then, $V_n^m(\bm{\theta^j})$ is the $n^{th}$ term of derivative operator of $U(\bm{\theta^j})$ for $m^{th}$ variable represented as,
\begin{widetext}
\begin{eqnarray*}
V_n^m(\bm{\theta})=c_n^m\prod_{l< m} \prod_k e^{-i\theta_l^j c_k^lP_k^l}\prod_{k < n}^{n - 1}e^{-i\theta_m^j c_k^mP_k^m}P_n^m\prod_{k \geq n}e^{-i\theta_m^j c_k^mP_k^m}\prod_{l> m} \prod_k e^{-i\theta_l^j c_k^lP_k^l}\label{V}.
\end{eqnarray*}
\end{widetext}
The advanced methods to calculate the derivatives have been proposed in \cite{2020arXiv201204429K} and \cite{2019arXiv190608728P}. The method that never requires ancillary qubits is proposed by Qunasys \cite{PhysRevResearch.2.013129}. However, this method uses a much larger number of gates compared to our method. On the other hand, the PSR is also omitted from our candidates of the method for the same reason \cite{2020arXiv201209265C}. The method we apply for calculating derivatives can be used for UCC ansatz. Therefore, we use the Hadamard-test-like method. The evaluation function for optimizing the norm of a tangent vector is,

\begin{equation}
F^{tv.}_j(\bm{\theta^j})=\sum_{m=0}\mid \frac{\partial E_j}{\partial \theta_m^j}\mid+E^{const}_j(\bm{\theta^j})+E^{def}_j(\bm{\theta^j})\label{Ftv}.
\end{equation}

\begin{figure*}[h]
 \centering
 \includegraphics[scale=0.7]{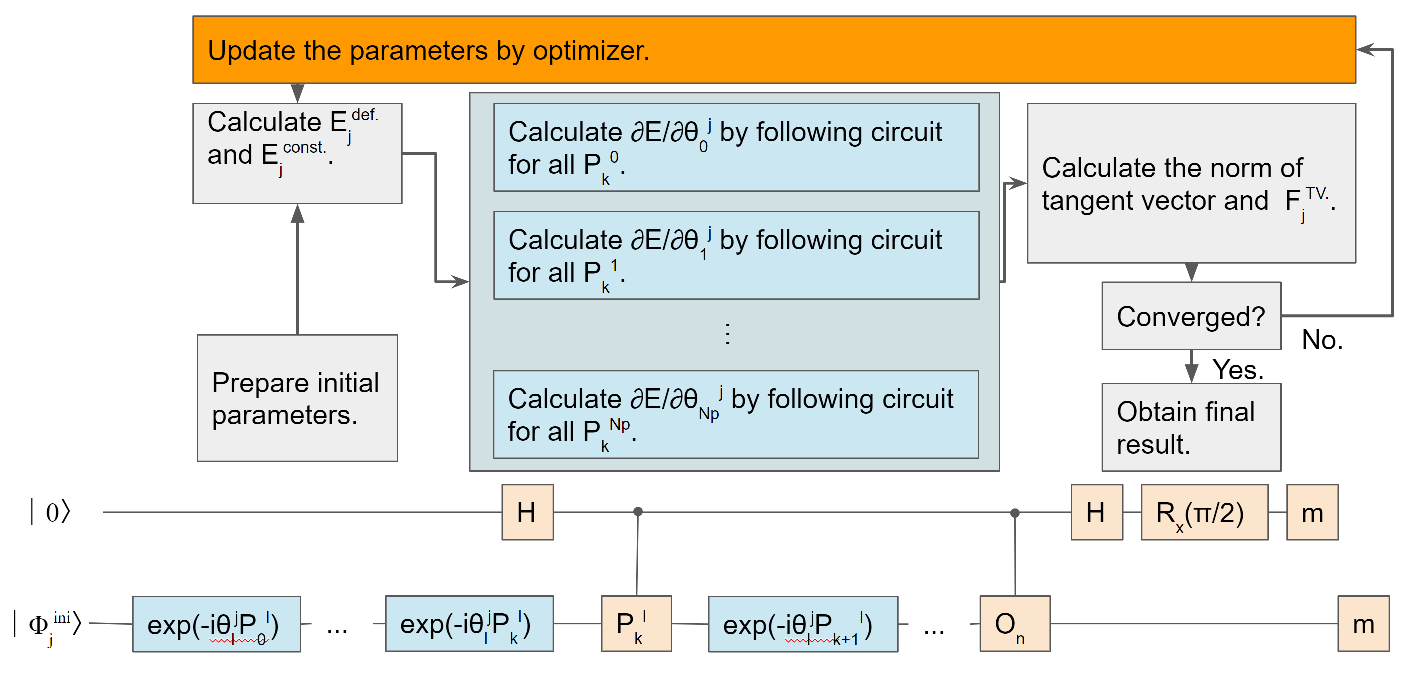}
 \caption{
 The flowchart of TVVQE and the quantum circuit to derive the derivative $\frac{\partial E_j }{\partial \theta_l^j}$.
 }\label{tvfig}
\end{figure*}

The detail of illustrated in Fig. \ref{tvfig}. The result of Variational Quantum Algorithms (VQAs) can be distinguished into three cases; (1) grobal minimum is derived, (2) local minimum is derived because the result is trapped by local minimum, (3) local minimum is derived because the minimum of evaluation function is smaller than grobal minimum of aimed state. The evaluation function $F^{tv.}$ has the minimum of zero. In contrast, the evaluation function of VQE is tend to be smaller than global minimum of aimed state due to smaller local minimums. TVVQE can avoid case (3). Optimized energy and states are calculated by the final variable vector. All above calculations are demonstrated by blueqat SDK \cite{blueqat}, a simulator of a quantum computer. All results of the quantum calculation are state vector (number of shots is infinity).

\section{Convergence for each system}\label{3.25}
\begin{figure}[h]
\includegraphics[scale=0.25]{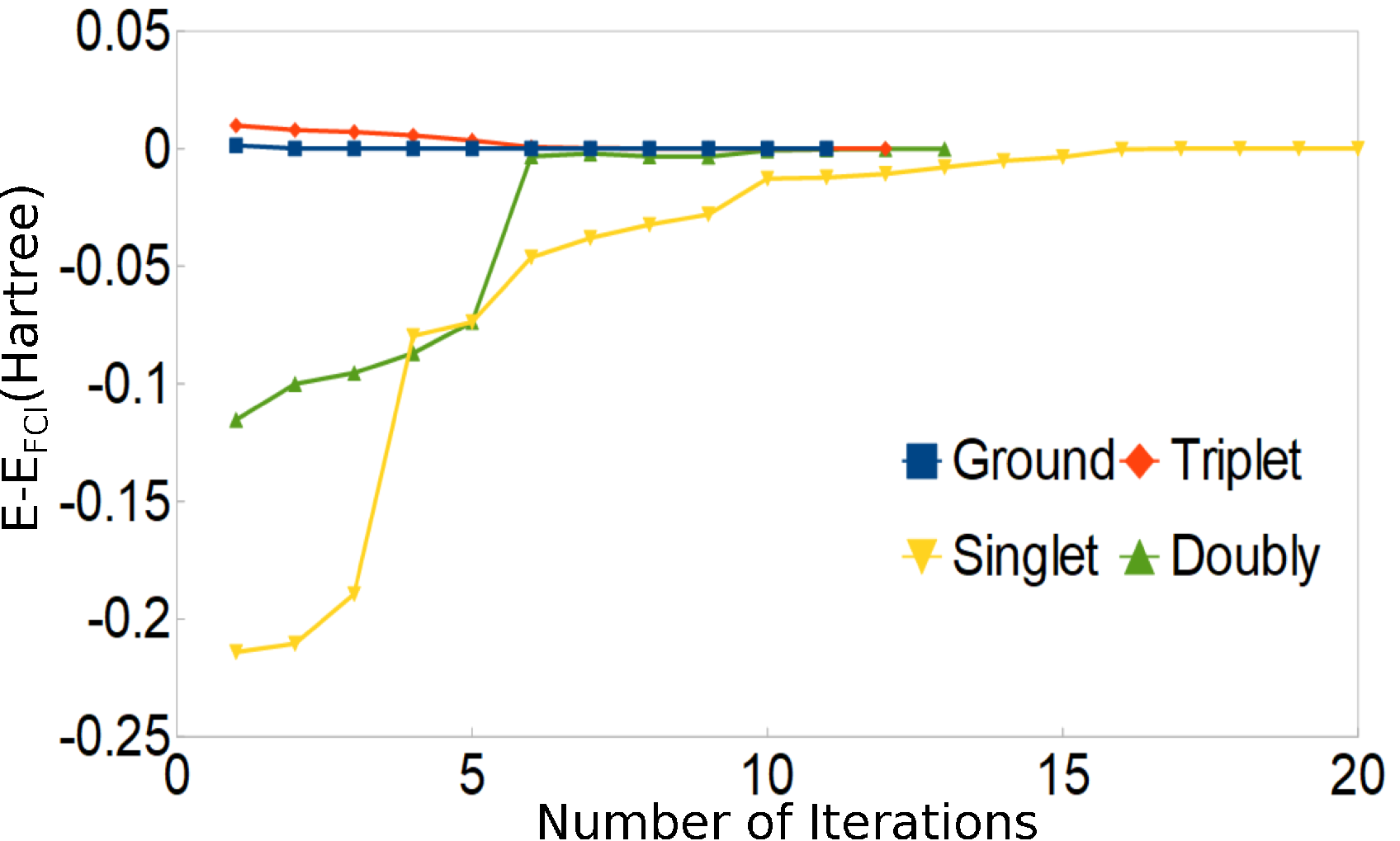}
\caption{
The number of iterations v.s. difference of calculated energy levels of ground, triplet, singlet, and doubly excited state from each exact value on hydrogen molecule calculated by TVVQE method.
}\label{Hoconv}
\end{figure}

In this section we describe the convergence of energy levels on a calculation of TVVQE for molecular Hamiltonian of a hydrogen molecule , rectangular Hubbard Hamiltonian sized 3$\times$ 1, the Hamiltonian of lithium hydride molecule, and the Hamiltonian of beryllium hydride molecule of only active orbitals \cite{2011H2305}. The orbitals to be calculated are chosen to be omit non molecular orbitals.  The Hamiltonian of the molecules is all prepared classically by STO-3G basis. The Hubbard Hamiltonian consists of a hopping term and a Coulomb term taking into account spins \cite{Jiang1424}, thus, it has 6 sites and 2 bonds. Hence this requires 6 qubits. The coefficient of the hopping term is $t=0.13 eV$ and the coefficient of the Coulomb term is $8t$, respectively. The method to optimize the evaluation function is Broyden-Fischer-Goldferb-Shanno (BFGS) method \cite{optimization}. This is quasi-Newtonian method that uses quasi-Hessian for optimization instead of ordinary Hessian. The number of iterations of the optimization process of tangent vector in TVVQE is 10 for all states. Firstly, we show the convergence of hydrogen molecules. As shown in Fig. \ref{Hoconv}, ground, triplet, singlet, and doubly excited state converge to an exact value. VQE with VQD cannot derive the energy levels of the doubly excited state because of low accuracy of lower states.  Although, according to the difference between calculated energy and the exact energy of STO-3G basis (log error), the accuracy of the energy levels of this state is beyond the chemical accuracy as shown in Table. \ref{convcomp}. Moreover, all states approach an exact value in the process of optimizing tangent vectors.

\begin{figure}[h]
\includegraphics[scale=0.25]{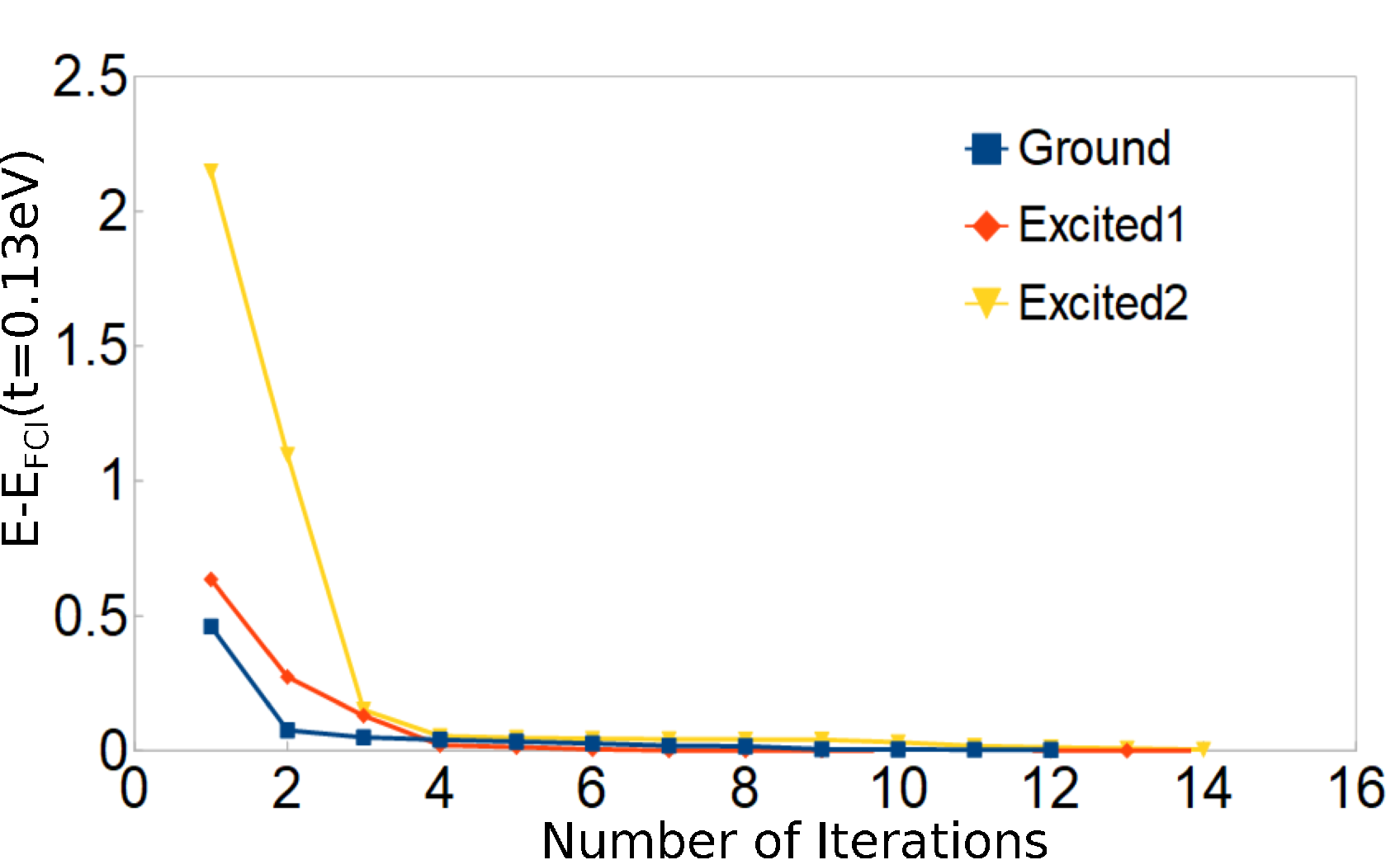}
\caption{
The number of iterations v.s. difference of calculated energy levels of the ground state and two excited states from each exact value on Hubbard model sized 3$\times$ 1 calculated by TVVQE method.
}\label{hubconv}
\end{figure}

\begin{figure}[h]
\includegraphics[scale= 0.12 ]{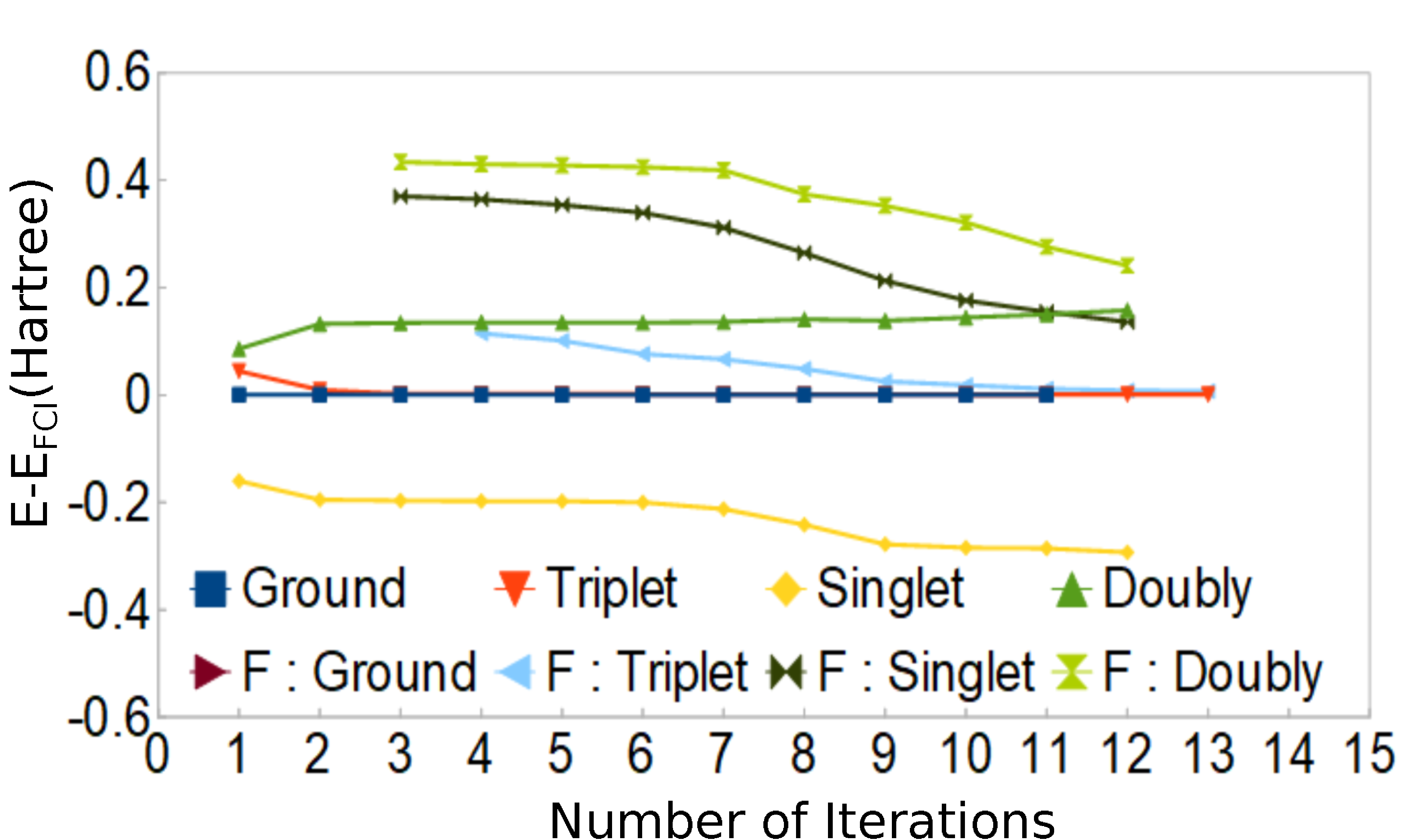}
\caption{
The number of iterations v.s. difference of calculated energy levels and the values of evaluation function of ground, triplet, singlet, and doubly excited state from each exact value on lithium hydride molecule calculated by TVVQE method. 
}\label{LiHconv}
\end{figure}

\begin{figure}[h]
\includegraphics[scale= 0.12 ]{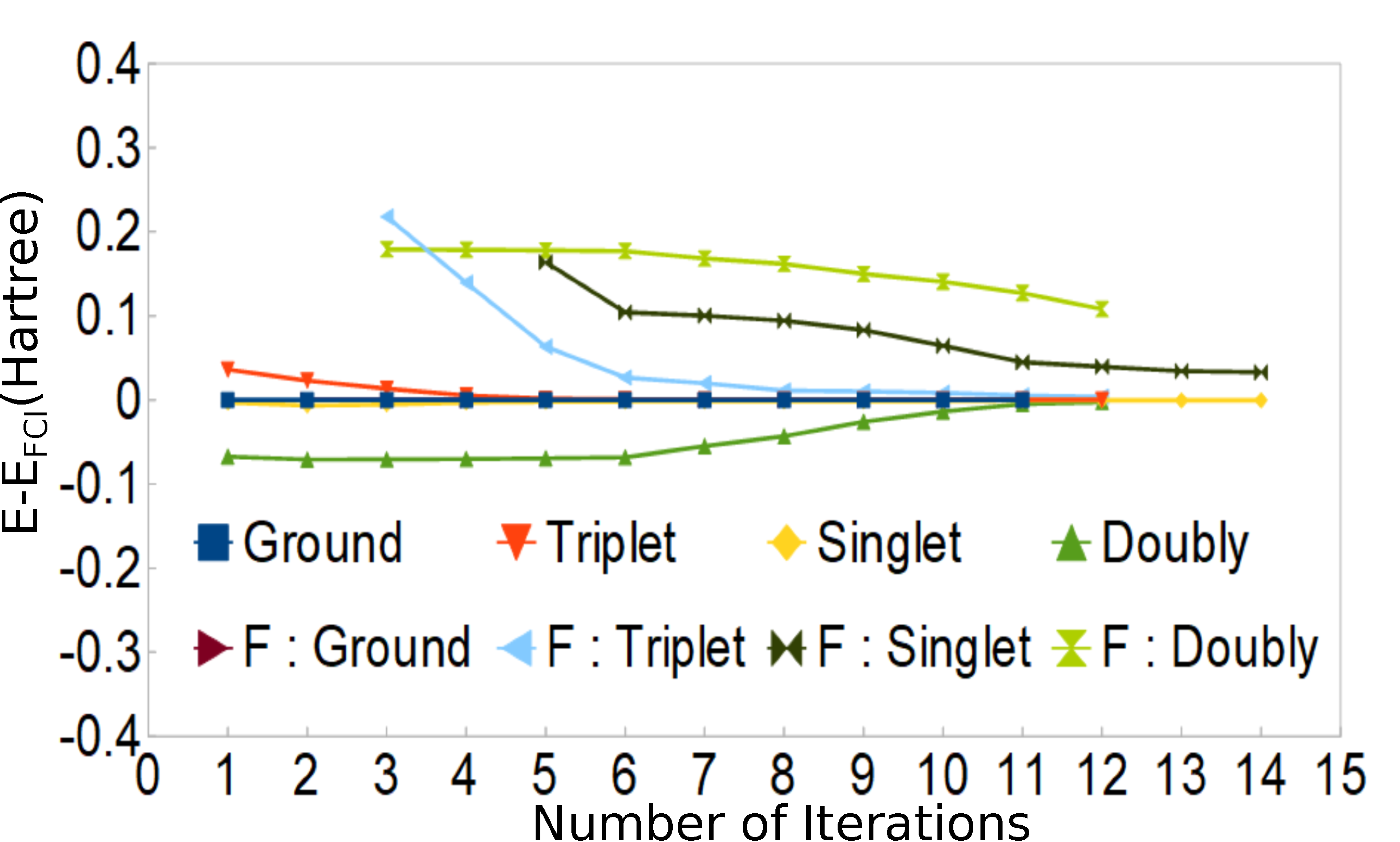}
\caption{
The number of iterations v.s. difference of calculated energy levels and the values of evaluation function of ground, triplet, singlet, and doubly excited state from each exact value on beryllium hydride molecule calculated by TVVQE method.
}\label{BeHconv}
\end{figure}

The energy levels of Hubbard Hamiltonian of size 3$\times$1 are also converged in high accuracy as shown in Fig. \ref{hubconv}. Even VQE with the VQD method converges in high accuracy, optimizing the tangent vector improves the accuracy of all states. The common logarithm of the absolute value of log errors of all states is below -2.5 as shown in Table. \ref{convcomp}. The energy levels of lithium hydride by reduced Hamiltonian have low accuracy as shown in Fig. \ref{LiHconv}. The energy of ground and triplet states only have high accuracy. Although, singlet and doubly excited states have low accuracy as shown in Table. \ref{convcomp} because the results are trapped by local minimums due to the little different from them. Singlet and doubly excited states are separated from the nearest local minimums by 0.293 and 0.1578 Hartree, respectively. Hence, the calculated singlet and doubly excited states are supposed to be trapped by the nearest local minimums. In fact, the difference between the calculated singlet state and the nearest local minimum is 0.0879 Hartree and doubly excited state and the nearest local minimum is 0.0482 Hartree, respectively, thus, they are closer than each exact value. However, the values of evaluation functions gradually closing to zero. Constraint terms $ E^{const}_j(\bm{\theta^j})+E^{def}_j(\bm{\theta^j}) $ may reach zero and exact energy are may be derived with more iterations.  On contrary, The energy levels of beryllium hydride by reduced Hamiltonian converge to an exact value as shown in Fig. \ref{BeHconv}. The ground , triplet and singlet states converged exact states same as others. doubly excited states approached to exact and accuracy is nearly chemical accuracy as shown in Table. \ref{convcomp}. The evaluation function have small values for singlet and doubly excited states. They have the room for improved.  

TVVQE is supposed to be subject to local minimums close to exact values. In contrast, the accuracy of results can be higher than that of VQE with VQD.

\begin{figure}[h]
\includegraphics[scale=0.25]{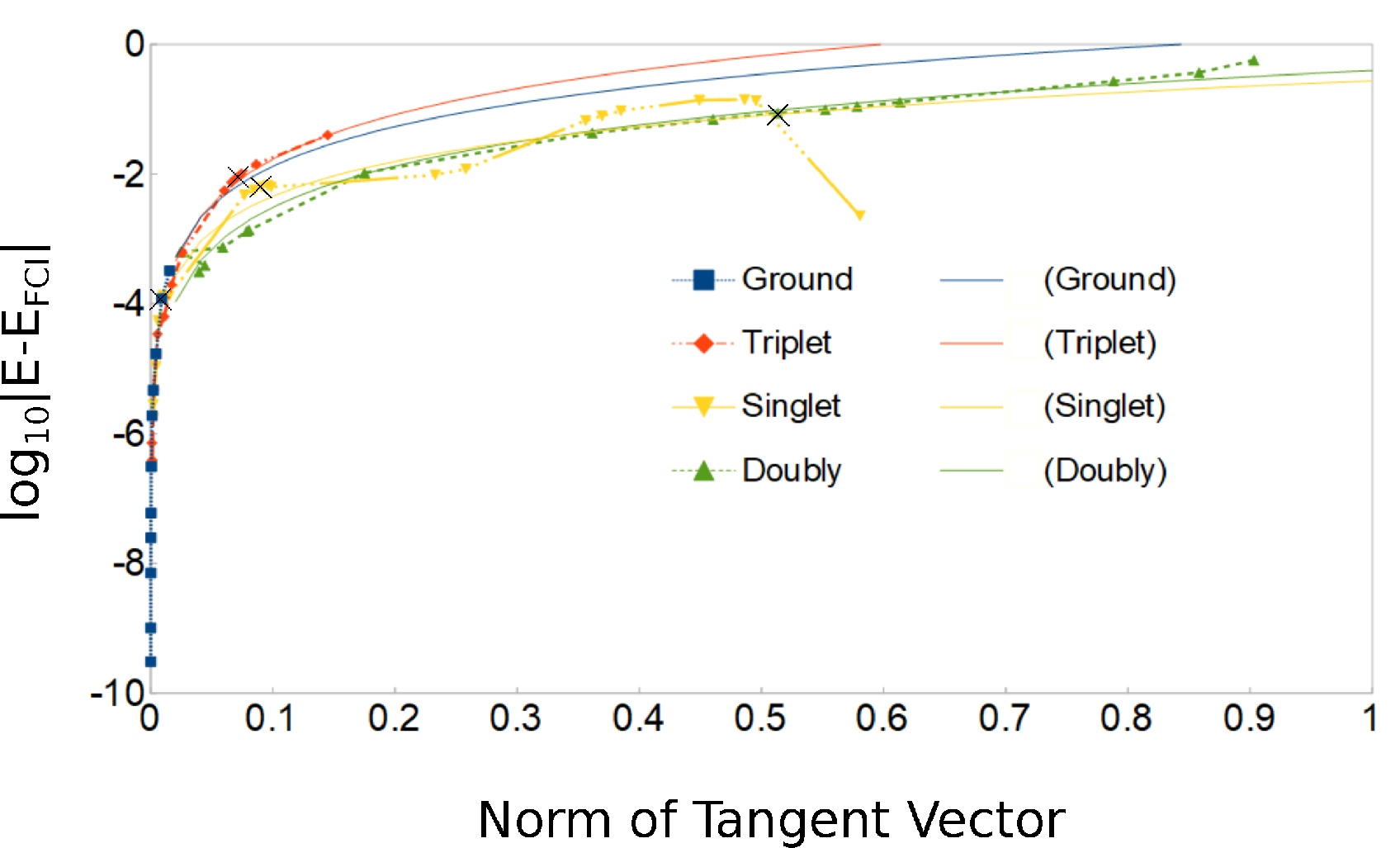}
\caption{
The norm of tangent vector v.s. log errors of energy levels of ground, triplet, singlet, and doubly excited state on hydrogen molecule calculated by TVVQE method. The start point of TVVQE is labeled by $\times$. Curved lines indicate the interpolated log errors by logarithm function by sampled data. TVVQE starts from 2$^{nd}$, 4$^{th}$, 12$^{th}$, and 7$^{th}$ points for ground, triplet, singlet, and doubly excited states, respectively.
}\label{Hocom}
\end{figure}

\begin{figure}[h]
\includegraphics[scale=0.25]{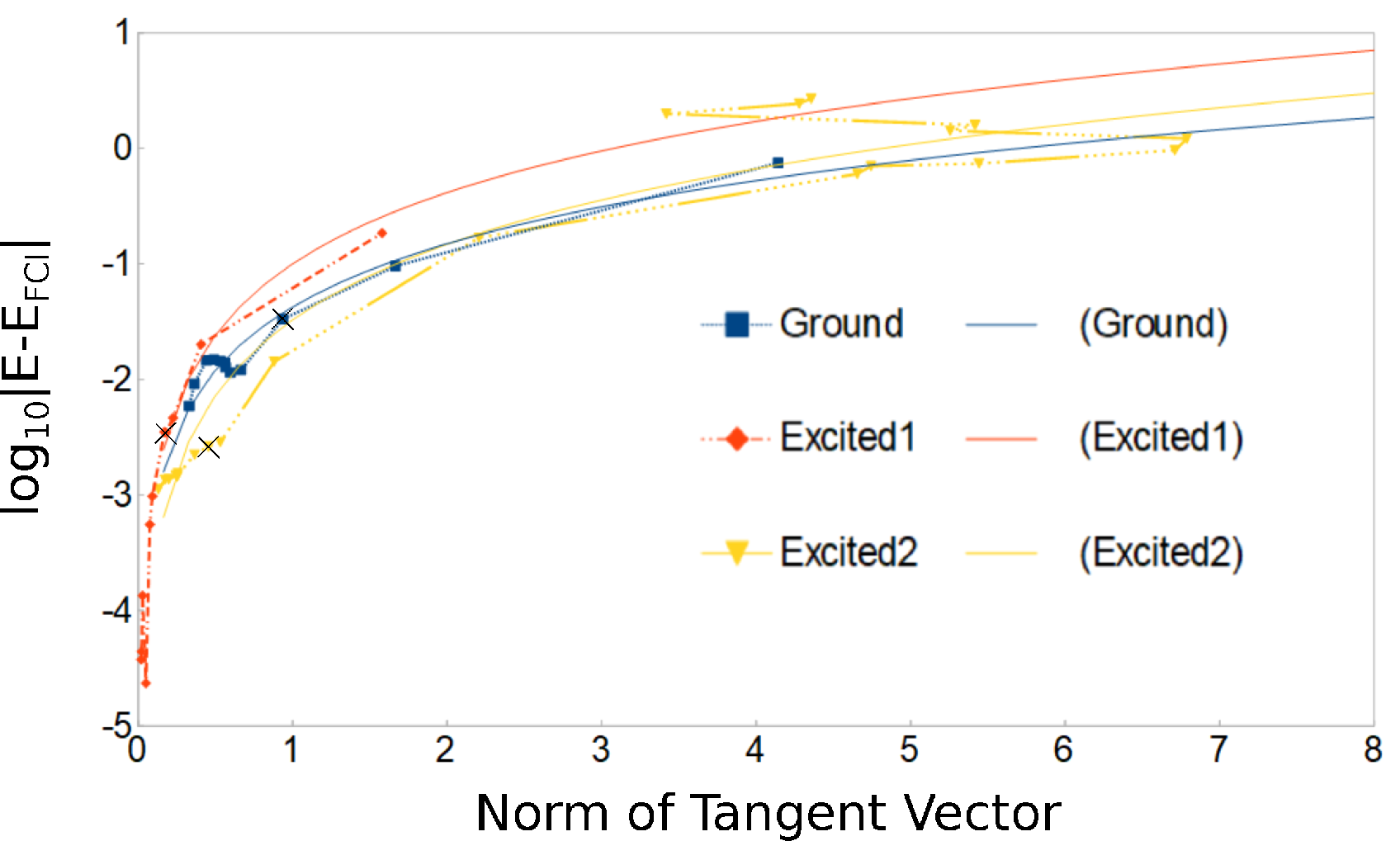}
\caption{
The norm of tangent vector v.s. log errors of energy levels of the ground state and two excited states on Hubbard model sized 3$\times$ 1 calculated by TVVQE method. TVVQE starts from 2$^{nd}$, 4$^{th}$ and 14$^{th}$ points for ground, excited1, and excited2 states, respectively.
}\label{hubcom}
\end{figure}

\begin{figure}[h]
\includegraphics[scale=0.25]{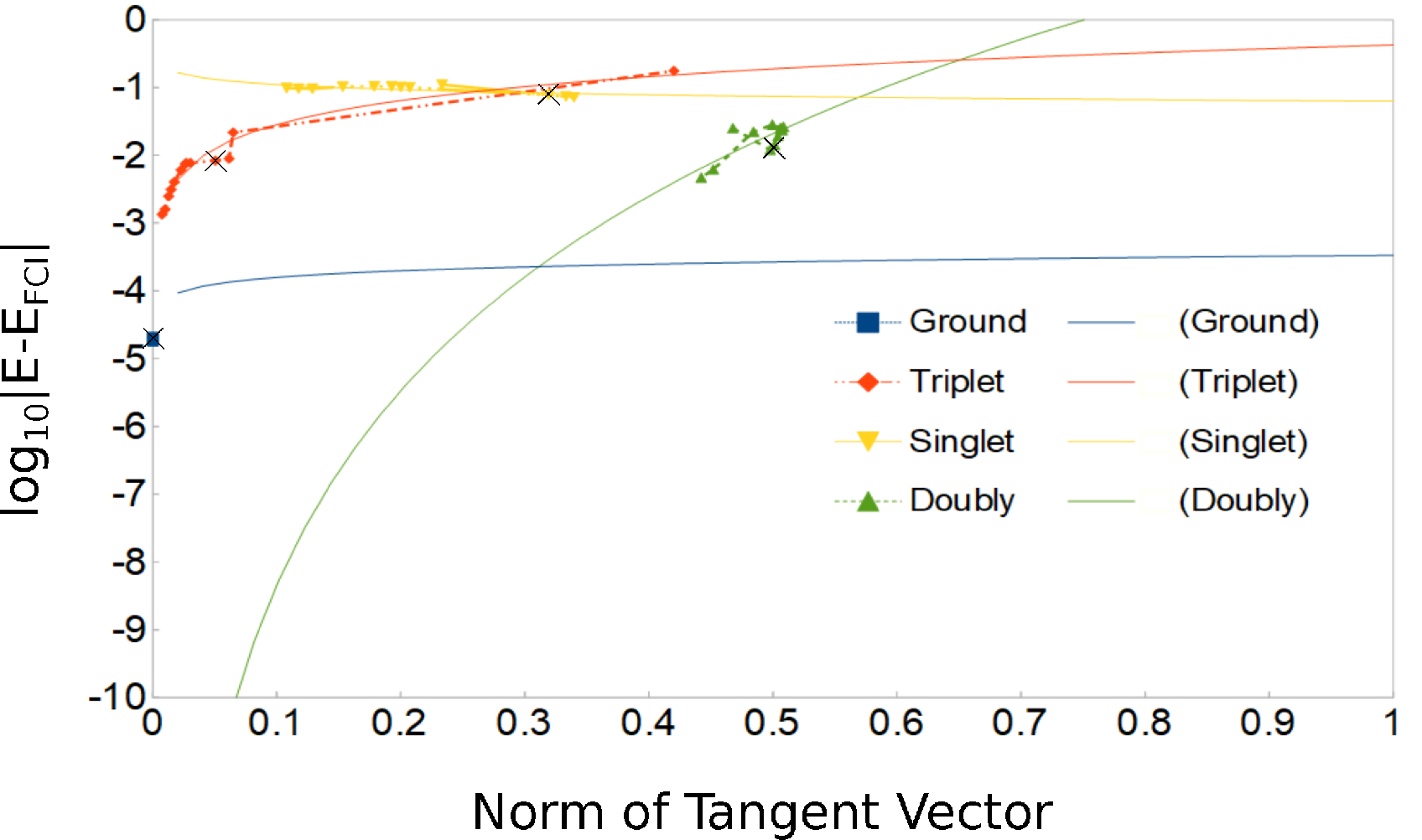}
\caption{
The norm of tangent vector v.s. log errors of energy levels of ground, triplet, singlet, and doubly excited state on lithium hydride molecule calculated by TVVQE method. TVVQE starts from 2$^{nd}$, 4$^{th}$, 3$^{rd}$, and 3$^{rd}$ points for ground, triplet, singlet, and doubly excited states, respectively.
}\label{LiHcom}
\end{figure}

\begin{figure}[h]
\includegraphics[scale= 0.25 ]{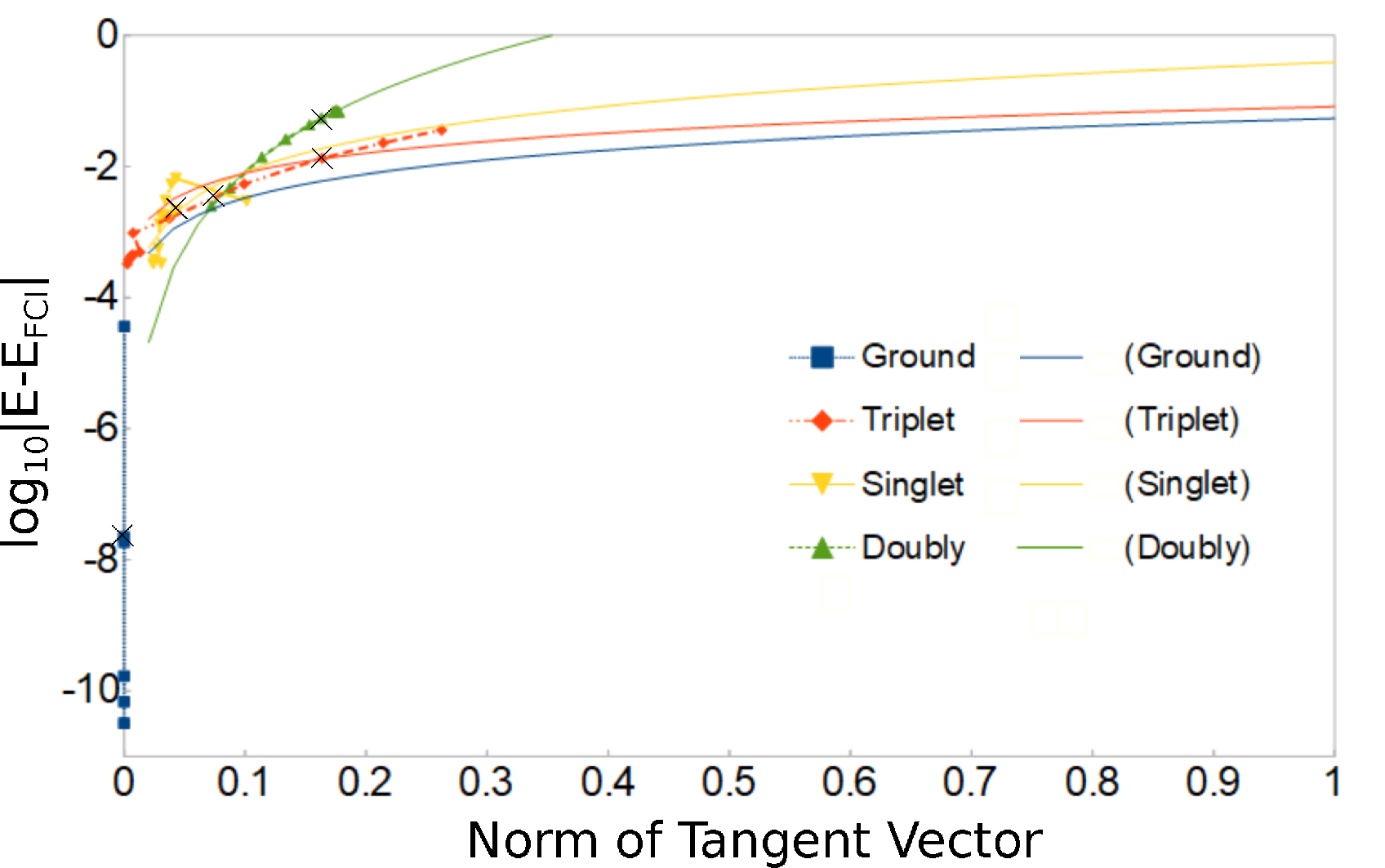}
\caption{
The norm of tangent vector v.s. log errors of energy levels of ground, triplet, singlet, and doubly excited state on beryllium hydride molecule calculated by TVVQE method. TVVQE starts from 2$^{nd}$, 3$^{rd}$, 5$^{th}$ and 3$^{rd}$ points for ground, triplet, singlet, and doubly excited states, respectively.
}\label{BeHcom}
\end{figure}

We also confirm the effect of optimization on tangent vectors. Fig. \ref{Hocom}, \ref{hubcom} , \ref{LiHcom} and \ref{BeHcom} are the plots of log error for the norm of the tangent vector of optimization on each state on hydrogen molecule, Hubbard model, and lithium hydride molecule, respectively. As for the hydrogen molecule and Hubbard model, log errors of all states are proportional to the norm of the logarithm of tangent vector as it gets small clearly. The exact values of excited states have large and peaky global minimums, thus, the initial value of the norm of tangent vector becomes larger as the energy level becomes higher. In contrast, the log errors of singlet and doubly excited states of lithium hydride molecule didn't converge to the exact values regardless of proportionality to the logarithm of the norm of the tangent vector. The singlet state is trapped by the nearest local minimums and the doubly excited state needs more iterations, respectively. The log error of the doubly excited state decreases in the process of TVVQE after the norm of the tangent vector increases. The log errors of proportional to the norm of the logarithm of tangent vector  same as hydrogen molecule.  It is supposed that the process of the TVVQE circumvents the value of the evaluation function from the local minimums because case (3) never occurs in TVVQE. The endeavor to make the initial state close to the exact value will contribute to the accuracy.

\begin{table}
\caption{
The table of log errors on converged energy levels and the number of iterations that start optimizing the norm of tangent vector of the calculation on hydrogen molecule, Hubbard lattice sized 3$\times $1, lithium hydride and beryllium hydride , respectively.
}\label{convcomp}
\begin{tabular}{cc|c|c|c|c} \hline \hline
System& &E$_0$&E$_1$&E$_2$&E$_3$ \\ \hline
H$_2$&log error&-12.4438&-6.8097&-5.3424&-3.8647 \\\cline{2-6}
&Start TV opt.&3&3&11&4 \\ \hline
Hubbard (3$\times$1)&log error&-2.5289&-5.5059&-2.3888&None. \\\cline{2-6}
&Start TV opt.&3&5&5&None. \\ \hline
LiH&log error&-3.3862&-3.044&-0.293&-0.8019 \\\cline{2-6}
&Start TV opt.&2&4&3&3 \\ \hline
BeH$_2$&&-10.4948&-3.4868&-3.4843&-2.5999 \\\cline{2-6}
&Start TV opt.&2&3&5&3 \\ \hline

\end{tabular}
\end{table}

\section{Comparison to other methods}\label{3.75}
\begin{figure*}[h]
\includegraphics[scale=0.475]{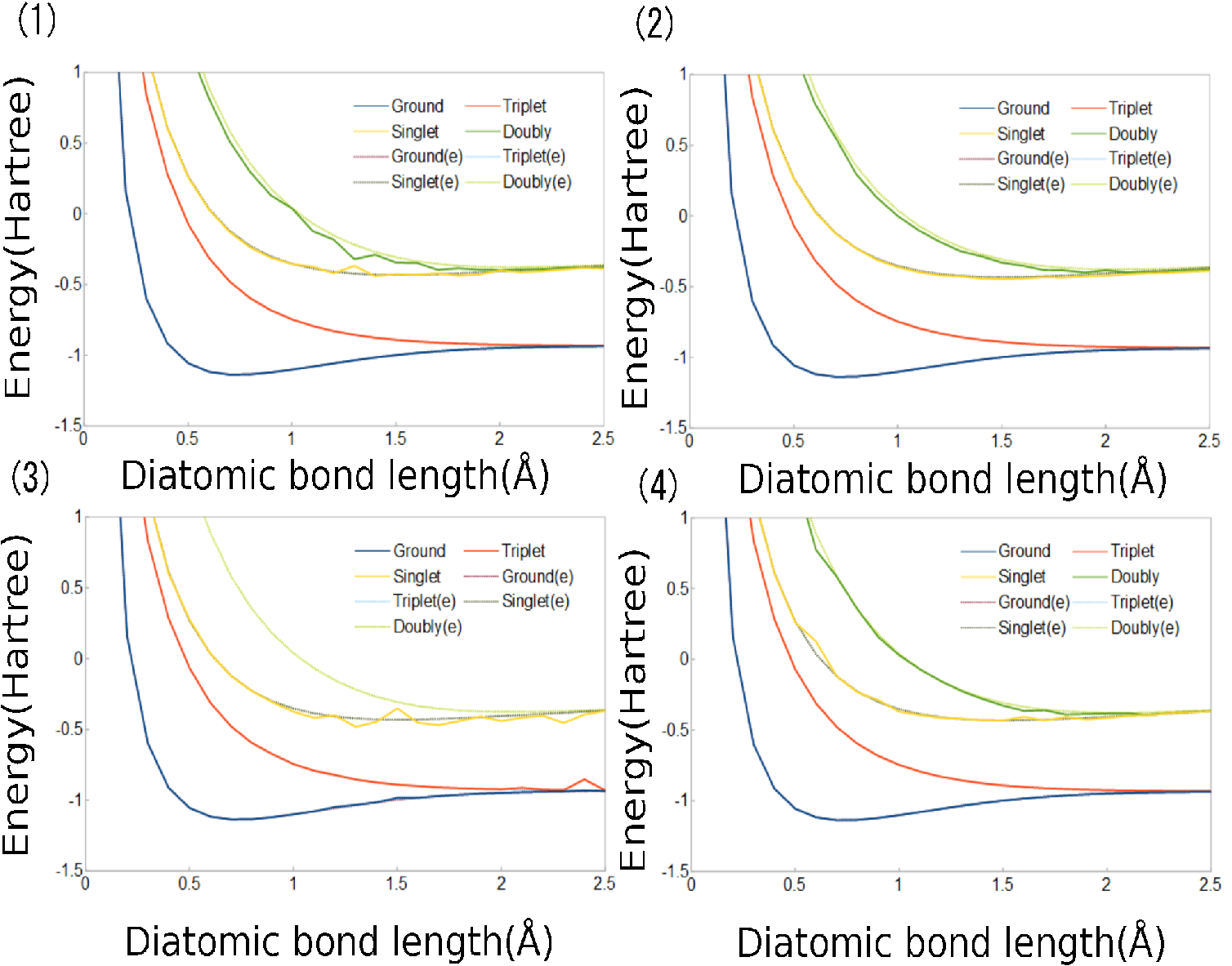}
\caption{
The diatomic bond length v.s. the energy levels of ground, triplet, singlet, and doubly excited states on hydrogen molecule, respectively calculated by (1)VQD, (2)Subspace-Search VQE (SSVQE) method, (3)Multiscale-Contracted VQE (MCVQE) method, and (4)Tangent-Vector VQE (TVVQE). The lines that have (e) in their suffix are exact values calculated by the Full-CI method.
}\label{Hcomp}
\end{figure*}

\begin{figure*}[h]
\includegraphics[scale=0.475]{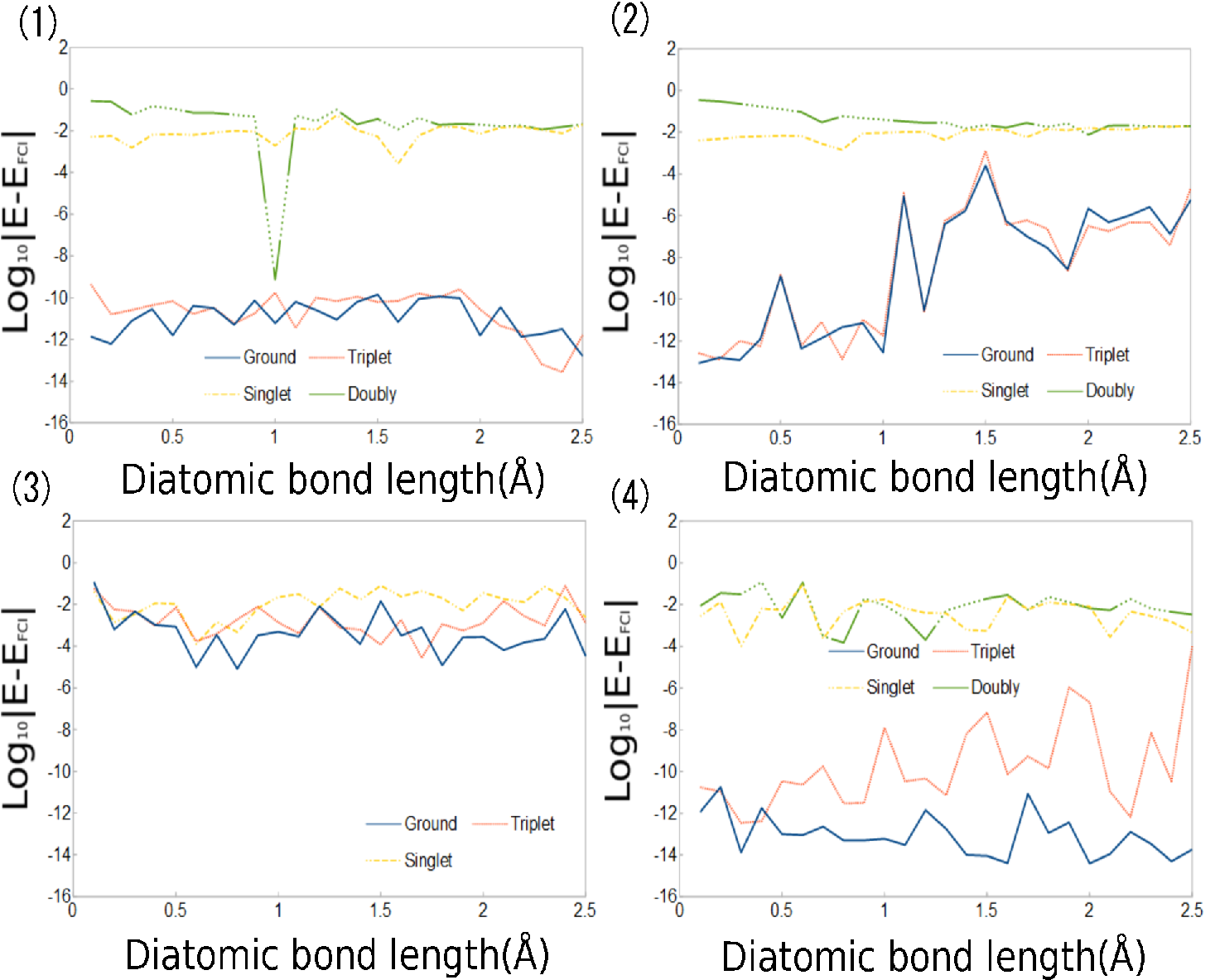}
\caption{
The diatomic bond length v.s. the log errors of ground, triplet, singlet, and doubly excited states on hydrogen molecule, respectively calculated by (1)VQD, (2)Subspace-Search VQE (SSVQE) method, (3)Multiscale-Contracted VQE (MCVQE) method, and (4)Tangent-Vector VQE (TVVQE). The lines that have (e) in their suffix are exact values calculated by the Full-CI method.
}\label{Hcompa}
\end{figure*}

In this section, we compare the result of the calculation on the ground, triplet, singlet, and double excited states of hydrogen molecule for diatomic bond length between hydrogen atoms from 0.1 to 2.5$(\AA)$ in 0.1 $(\AA)$ pitch using VQE with VQD method, SSVQE and MCVQE methods. The optimizer for all of the methods is the BFGS method. The number of iterations of (1)VQE with VQD is 22 and (2)SSVQE and (3)MCVQE method are 50, respectively. The details of each method are described in Appendix. The number of iterations of the optimization process of tangent vector in TVVQE is 3 for ground and 2 for all excited states. We show the result of the calculation on energy levels of these states and log errors using (1)VQE with VQD, (2)SSVQE, (3)MCVQE, and (4)TVVQE method on Fig. \ref{Hcomp} and \ref{Hcompa}, respectively. The calculated energy levels of (1)VQE with VQD match to exact values on many points for all states except the doubly excited state. Also shown in Table. \ref{Hcompl}, triplet, and doubly excited states have the highest accuracy on average in all methods. The calculated energy levels of (2)SSVQE method are close to exact values on almost all points superficially. However, the average log errors for diatomic bond length r are larger than those of (1) and (4) as shown in Table. \ref{Hcompl}. The calculated energy levels of (3)MCVQE method are the least accurate for ground and triplet states as shown in Table. \ref{Hcompl}. The energy levels of (4)TVVQE match to exact values on almost all points except the doubly excited state same as (1). However, more points of a doubly excited state match to exact values than (1). The averages of log error of ground and singlet states are the smallest in all methods as shown in Table. \ref{Hcompl}.
The number of iterations of the optimization process of the tangent vector is the least number that the improvement of accuracy of TVVQE occurs. The required least number of iterations to calculate the energy levels accurately more than chemical accuracy is revealed to be smaller than the ordinary VQE method as predicted from the result in the paper \cite{2017arXiv170102691R}.

\begin{table}[h]
\caption{
The average of log error of ground, triplet, singlet and doubly excited state on hydrogen molecule for diatomic bond length r of (1)VQD, (2)Subspace-Search VQE (SSVQE) method, (3)Multiscale-Contracted VQE (MCVQE) method, and (4)Tangent-Vector VQE (TVVQE).
}\label{Hcompl}
\begin{tabular}{c|c|c|c|c}\hline\hline
Method&Ground&Triplet&Singlet&Doubly \\ \hline
VQD&-10.9728&-10.7078&-2.1267&-1.6978 \\
SSVQE&-8.6116&-8.5508&-2.0689&-1.4055 \\
MCVQE&-3.3662&-2.7753&-1.9895&None. \\
TV-VQE&-12.8133&-9.0578&-2.1978&-1.6414 \\ \hline
\end{tabular}
\end{table}

\section{Conclusion}\label{4}
In this work, it is confirmed that TVVQE can calculate the energy levels and states more accurately than other conventional VQE methods for Hubbard model, hydrogen, lithium hydride Hamiltonians. This means that optimizing the norm of the tangent vector of trial energy contributes to the accuracy. Moreover, the accuracy of the calculation by TVVQE is better than other conventional methods even when the number of iterations is limited to 2 times. This means that the accuracy will be higher as the number of iterations increases a little. However, the time for calculation of TVVQE is five times longer than that of VQE with VQD on average because each process of TVVQE requires far more times than ordinary VQE due to the calculation times of simulations of Hadamard test. It can be decreased by calculating derivatives in parallel by a sufficient number of the quantum circuit for the number of variable parameters. Adaptive VQE can be improved the time for calculations by only decreasing operation times. TVVQE can be improved also by parallel computing.  Hence, TVVQE has the potential to derive the energy levels more accurately and quickly than other VQE algorithms.

\newpage

\bibliographystyle{apsrev4-2}
\bibliography{temp6}
\newpage
\clearpage
\appendix{Appendix : The details of each method}\label{A}

In this section, we describe the details of our methods in Section \ref{3.75}. The deflation term of i$^{th}$ state on VQE and TVVQE methods is as follows,

\begin{eqnarray}
E_{i}^{def}&=&((af+b(1-f)) \\ \nonumber
&\times& (\sum_{j<i}(exp(r-0.25r_d)+1)^{-1} \\ \nonumber
&\times & \mid\langle\Phi_j\mid\Phi_i\rangle\mid{^2}\\ \nonumber
&+&(1-(exp(r-0.25r_d)+1)^{-1}) \\ \nonumber
& \times & f(\mid\langle\Phi_j\mid\Phi_i\rangle\mid{^2})).\label{defmisc}
\end{eqnarray}

The $a$, $b$, and $f=(exp(\alpha(r-r_d))+1)^{-1}$ are functions for two constants, and Fermi-Dirac distribution for $r$, respectively. Furthermore, $r_d$ is a constant bond length in molecular hydrogen. $f(\mid\langle\Phi_j\mid\Phi_i\rangle\mid{^2})$ is a quadratic function of the absolute value of the inner product between the i- and j-states to accurately derive the degenerate states. In this paper, $a=1.0$, $\alpha=100$ and $f(\mid\langle\Phi_j\mid\Phi_i\rangle\mid{^2})=(1+2(\sqrt{5}+1))r^4/r_d^4E_p(r)/4\mid\langle\Phi_j \mid\Phi_i\rangle\mid{^4}+2(\sqrt{5}+1)r^4/r_d^4E_p(r)/4\mid\langle\Phi_j\mid\Phi_i\rangle\mid{^2}$, where $E_p(r)$ is the energy of one lower state of molecular hydrogen at bond length $r$.

Deflation term of i$^{th}$ on SSVQE and MCVQE is the only inner product of states. It is the product of inner product of i$^{th}$ and j$^{th}$ states and variable $\lambda_i + \lambda_j$. Then, $\lambda_i$ is $2(N - i)/(N^2-N)$ for the number of states $N$. Initial states of ground, triplet, singlet and doubly excited states are $\mid 1000 \rangle, \mid 1100 \rangle, \mid 0110 \rangle$ and $\mid 0010 \rangle$, respectively. The process to make the CIS state is performed after SSVQE in MCVQE. The process of SSVQE in the MCVQE method is the same as normal SSVQE that optimizes ground, triplet, and singlet states, respectively.

\end{document}